\documentclass[12pt]{iopart}

\usepackage{graphicx}

\begin{document}

\title[Saturation of the filamentation instability]
{The saturation of the electron beam filamentation instability by 
the self-generated magnetic field and magnetic pressure gradient-driven 
electric field}

\author{M E Dieckmann\textsuperscript{1,2,3}, G Rowlands\textsuperscript{4},
I Kourakis\textsuperscript{1} and M Borghesi\textsuperscript{1}}

\address{1 Centre for Plasma Physics, Queen's University Belfast, Belfast
BT7 1NN, U K}
\address{2 Theoretische Physik IV, Ruhr-University Bochum, 44780 Bochum,
Germany}
\address{3 Department of Science and Technology (ITN), Link\"oping 
University, Campus Norrk\"oping, 60174 Norrk\"oping, Sweden}
\address{4 Physics Department, Warwick University, Coventry CV4 7AL, U K}
\ead{Mark.E.Dieckmann@itn.liu.se}

\begin{abstract}
Two counter-propagating cool and equally dense electron beams are modelled
with particle-in-cell (PIC) simulations. The electron beam filamentation
instability is examined in one spatial dimension. The box length resolves
one pair of current filaments. A small, a medium-sized and a large filament
are considered and compared. The magnetic field amplitude at the saturation
time of the filamentation instability is proportional to the filament 
size. It is demonstrated, that the force on the electrons imposed by the 
electrostatic field, which develops during the nonlinear stage of the 
instability, oscillates around a mean value that equals the magnetic 
pressure gradient force. The forces acting on the electrons due to 
the electrostatic and the magnetic field have a similar strength. The 
electrostatic field reduces the confining force close to the stable 
equilibrium of each filament and increases it farther away. The confining 
potential is not sinusoidal, as assumed by the magnetic trapping model, 
and it permits an overlap of current filaments (plasmons) with an opposite 
flow direction. The scaling of the saturation amplitude of the magnetic 
field with the filament size observed here thus differs from that expected 
from the magnetic trapping model. The latter nevertheless gives a good 
estimate for the magnetic saturation amplitude. The increase of the peak 
electrostatic and magnetic field amplitudes with the filament size implies, 
that the electrons heat up more and that the spatial modulation of their 
mean speed along the beam flow direction increases with the filament size. 
\end{abstract}


\maketitle

\section{Introduction}
The electron beam filamentation instability (FI) generates magnetic 
fields in energetic astrophysical and solar flare plasmas 
\cite{Gallant,Petri,Karlicky,Stock,Niemiec}. The FI is also relevant 
for inertial confinement fusion (ICF) \cite{Tabak,Macchi,Macchi2} 
and laboratory astrophysics \cite{Woolsey} experiments, as well as 
for particle accelerators \cite{Dav1,Dav2}. The FI is driven by 
counter-propagating electron beams and it feeds on their mean flow energy.
This contrasts the Weibel instability, which grows magnetic fields at the 
expense of a thermal anisotropy \cite{Weibel,Yoon,Davidson}. 
The Weibel instability and the FI can be combined to form cumulative 
instabilities \cite{Lazar}. The FI becomes important, if the beam speeds 
are at least mildly relativistic.

The growth and saturation of the FI can be modelled with particle-in-cell 
(PIC) or Vlasov codes. The FI has been investigated with a one-dimensional 
(1D) Vlasov code \cite{DavidsonO,Califano} and with a two-dimensional (2D) 
PIC code \cite{Pukhov}, taking into account the ion dynamics. The impact 
of a flow-aligned magnetic field has been examined in Ref. \cite{Stockem}. 
PIC simulation studies have addressed the statistical properties of the 
FI in 1D \cite{Rowlands} and in 2D \cite{Silva,Dieckmann}. The 
counterstreaming electron instability has also been examined with 3D PIC 
simulations \cite{Sakai}.

The probably simplest and thus widely researched plasma configuration that 
gives rise to the FI consists of equally dense and equally warm electron 
beams that have a Maxwellian velocity distribution. Their thermal velocity 
spread $v_{th}$ in the rest frame of the respective beam is the same in 
all directions. This system can be considered in a simulation reference 
frame, in which both beams move into opposite directions at the 
non-relativistic speed modulus $v_b$ and with $v_{th} \ll v_b$. The FI 
can be isolated by selecting a 1D or 2D simulation box, that is oriented 
orthogonally to the beam velocity vector $\mathbf{v}_b$. The electron 
velocities must be resolved in the simulation direction or plane and along 
the beam direction. The FI competes in reality with the two-stream modes 
and it can probably not be observed experimentally in an isolated form, 
even if the equal beam densities favor the FI over the two-stream 
instability \cite{Bret}. However, the gained insight into the development 
of the isolated FI will improve the understanding of systems, in which the 
FI interplays with other instabilities. 
 
The linear and nonlinear evolution of the FI driven by counter-propagating
identical electron beams is qualitatively understood, at least in one 
spatial dimension where the filament mergers are suppressed \cite{Lampe}. 
The Refs. \cite{Davidson,Lampe,Califano,Stockem,Rowlands} have provided 
an insight into its linear and nonlinear development, which can be 
summarized as follows. The FI triggers the aperiodic growth of waves out
of an initial perturbation (noise) with the wavevectors $\mathbf{k} \perp 
\mathbf{v}_b$ over a wide band of $k=|\mathbf{k}|$, where the wavenumbers 
$k$ are of the order of the inverse electron skin depth. The electrons are 
deflected by the magnetic field perturbation, and electrons moving in
opposite directions separate in space. The net current of these flow
channels amplifies the initial perturbation and, thus, the tendency to
form current channels. The magnetic field amplitude grows exponentially. 
Magnetic trapping has been identified as a possible saturation mechanism 
\cite{Davidson}. It has also been proposed \cite{Califano} that the 
electrostatic fields, which grow during the nonlinear evolution of the 
FI, may be important for the saturation. These electrostatic fields have 
been related to the magnetic pressure gradient \cite{Stockem,Rowlands}. 
However, it has not yet been demonstrated quantitatively that the 
electrostatic fields during the quasi-linear evolution of the FI do 
originate from the magnetic pressure gradient. A direct comparison of 
the relative importance of the electric and magnetic fields for the 
nonlinear saturation of the FI is also lacking and this is an objective 
of this paper.

The length of the 1D simulation box with periodic boundaries can be 
selected such, that only one wave period grows. This FI evolves like that 
in a much larger box \cite{Rowlands} and the saturation mechanism should 
thus be the same. We employ here a simulation box that resolves a single 
spatial period of the growing wave and we can thus compare our results to 
previous work \cite{Califano}. We employ plasma parameters that are 
identical to those in Ref. \cite{Rowlands} and resort to the distribution 
of the filament sizes, which is computed in that paper. We perform three 
simulations, in which we vary the box length. The spectrum of unstable 
wavenumbers of the FI is bounded at low and large wavenumbers, the latter 
by thermal effects \cite{Califano}. The bounded $k$-spectrum implies in 
turn a maximum and a minimum filament size. We model a filament size close 
to the minimum value, one close to the maximum value and one, that is close 
to the size with the maximum probability. We compare the properties of the 
filaments.

This paper is structured as follows. Section 2 discusses the PIC code 
and the initial conditions. The results are presented in the section 3.  
All simulations demonstrate that the electrons are heated up orthogonally 
to $\mathbf{v}_b$, in line with previous simulations \cite{Rowlands}. The
heating is achieved by the simultaneous interaction of the electrons with 
the quasi-static magnetic field and the oscillatory electrostatic fields. 
The heating is much stronger for the larger filaments than for the small 
one. The magnetic field amplitude reaches a value that is in reasonable 
agreement with the one expected from the magnetic trapping mechanism. It 
does, however, not scale correctly with the filament size. A reason is
that the electrostatic field modifies the potential. The force excerted by 
the electrostatic field in the simulation of the least turbulent small 
filament oscillates around a mean value that equals the magnetic pressure 
gradient force, confirming experimentally their connection. The fields
of the two larger filaments show the same spatial correlation. 
The heated electrons cannot be confined by the electromagnetic field but 
a cooler, dense electron population remains localized, forming a plasmon.
This plasmon can propagate at a sizeable fraction of the initial electron 
thermal speed, which contrasts the non-propagating filamentation modes out
of which the plasmon forms. The plasmon maintains the net current along 
$\mathbf{v}_b$ and, thus, the magnetic field. The slow extraordinary mode 
is pumped in the two large simulation boxes. However, the resulting growth 
of the discrete wave spectrum observed here, which has been reported first
by Ref. \cite{Califano}, is a finite box effect. The results are discussed 
in more detail in section 4.

\section{Solved equations and initial conditions}

The particle-in-cell simulation method is detailed in Ref. \cite{Dawson}. 
Our code is based on the numerical scheme proposed by \cite{Eastwood}. A 
PIC code approximates a phase space fluid by an ensemble of computational 
particles (CPs). The CPs can have a mass $m_{cp}$ and charge $q_{cp}$ that 
differs from the physical particle it represents, but the charge-to-mass 
ratio must be preserved. The equations, which the PIC code is solving, are
\begin{eqnarray}
\nabla \times \mathbf{E} = -\frac{\partial \mathbf{B}}{\partial t} \, , 
\, \, \nabla \times \mathbf{B} = \mu_0 \mathbf{J} + \mu_0 \epsilon_0 
\frac{\partial \mathbf{E}}{\partial t}, \\
\nabla \cdot \mathbf{E} = \rho / \epsilon_0 , \, \, \, \nabla \cdot
\mathbf{B} = 0, \\
\frac{d}{dt} \mathbf{p}_{cp} = q_{cp} \left ( \mathbf{E}[\mathbf{x}_{cp}] 
+ \mathbf{v}_{cp} \times \mathbf{B}[\mathbf{x}_{cp}] \right ) \, , \, \, 
\frac{d}{dt} \mathbf{x}_{cp} = \mathbf{v}_{cp},
\end{eqnarray}
with $\mathbf{p}_{cp} = m_{cp} \Gamma_{cp} \mathbf{v}_{cp}$. The 
currents $\mathbf{j}_{cp}=q_{cp} \mathbf{v}_{cp}$ of each CP are 
interpolated to the grid. The summation over all CPs gives the macroscopic 
current $\mathbf{J}$, which is defined on the grid. The $\mathbf{J}$ 
updates the $\mathbf{E},\mathbf{B}$ fields through Eq.(1). The updated 
fields are interpolated to the position of each CP and update its
position $\mathbf{x}_{cp}$ and $\mathbf{p}_{cp}$ through Eq.(3). This 
scheme is repeated over $N_s$ time increments $\Delta_t$. 

Two equally dense beams of electrons with $q_{cp}/m_{cp}=-e/m_e$ move 
along the $\mathbf{z}$-direction. Beam 1 has the mean speed $\mathbf{v}_{b1} 
= v_b \mathbf{z}$ and the beam 2 has $\mathbf{v}_{b2} = -v_b \mathbf{z}$ 
with $v_b / c = 0.3$. Both beams have a Maxwellian velocity distribution 
in their respective rest frame with a thermal speed $v_{th}={(k_b T / 
m_e)}^{0.5}$ of $v_b / v_{th} = 18$. Both beams are spatially uniform. 
The 1D simulation box with its periodic boundary conditions is aligned 
with the $\mathbf{x}$-direction. We thus denote positions by the scalar 
$x$. The plasma frequency of each beam with the number density $n_e$ is 
$\omega_p = {(e^2 n_e / m_e \epsilon_0)}^{0.5}$. The total plasma frequency 
$\Omega_p =\sqrt{2}\omega_p$. The electric and magnetic fields are
normalized to $\mathbf{E}_N = e\mathbf{E}/c m_e \Omega_p$ and $\mathbf{B}_N 
= e\mathbf{B} /m_e \Omega_p$ and the current to $\mathbf{J}_N = \mathbf{J} 
/ 2 n_e e c$. The physical position and time are normalized as $x_N = x/ 
\lambda_s$ with the electron skin depth $\lambda_s = c / \Omega_p$ and 
$t_N = t \Omega_p$. We drop the indices $N$ and $x,t,\mathbf{E},\mathbf{B},
\mathbf{J}$ are specified in normalized units.

The x-direction is resolved in the three simulations by $N_g = 500$ grid 
cells with the length $\Delta_x$. The phase space distributions 
$f_1(x,\mathbf{p})$ of beam 1 and $f_2(x,\mathbf{p})$ of beam 2 are each
sampled by $N_p = 6.05 \cdot 10^7$ CPs. The total phase space density is 
defined as $f(x,\mathbf{p})=f_1(x,\mathbf{p})+f_2(x,\mathbf{p})$. The box 
length of run 1 is $L_1=2.8$, that of run 2 is $L_2=2$ and that of run 3 
is $L_3=0.89$. We use $L_1,L_2,L_3$ to label the simulation runs. According 
to the size distribution in Ref. \cite{Rowlands}, we may expect that a 
single wave period grows in each box. The simulation $L_3$ captures the 
smallest and $L_1$ the largest filament that can grow with a significant 
probability. The simulation $L_2$ corresponds to a filament size close to 
that, which grows with the maximum probability. Multiple small filaments 
of a size $\approx L_3$ could grow in the larger simulation boxes. This is 
initially observed in the largest box $L_1$, but the smaller filaments 
merge to give one large one.
 
\section{Simulation results}

The selected beam velocity vector $\mathbf{v}_b \parallel \mathbf{z}$ 
and the simulation box orientation imply, that the electrons of both 
beams and their micro-currents are re-distributed by the FI only along 
$x$. The charge- and current-neutral plasma is transformed into one with 
$J_z (x) \neq 0$. According to Eq.(1) a growth of $J_z(x)$ is coupled 
only into the $E_z (x)$ and $B_y (x)$ field components. This is, because 
the gradients along the $y,z$-direction vanish in the 1D geometry. Ampere's 
law simplifies to $dB_y / dx= J_z + \partial E_z /\partial t$. A $J_z \propto 
\sin{(kx)}$ gives a $E_z \propto \sin{(kx)}$ and $B_y \propto -\cos{(kx)}$ 
so that $E_z$ and $B_y$ will have a phase shift of $90^\circ$. The electron 
re-distribution leads to a space charge and thus to a $E_x (x) \neq 0$. 

\begin{figure}
\begin{center}
\includegraphics[width=8.2cm]{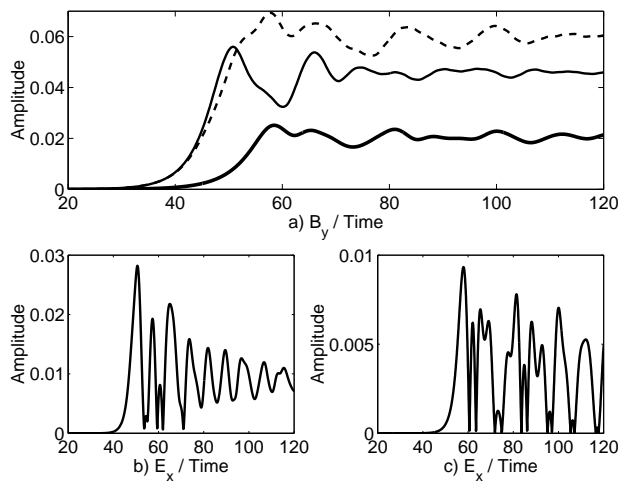}
\caption{Field amplitudes: (a) shows the $B_y (k_1,t)$ for the box $L_3$ 
(thick solid line), the box $L_2$ (thin solid line) and the box $L_1$ 
(dashed line). The amplitudes grow exponentially, they saturate and
oscillate around an equilibrium value. This value increases with the
box size. (b) shows $E_x (k_2,t)$ for the box $L_2$ and (c) that for 
the box $L_3$. The $E_x$ component evolves towards an equilibrium value
for the box $L_1$ (not shown) and $L_2$ and $E_x$ remains oscillatory for 
the box $L_3$.\label{fig1}}
\end{center}
\end{figure}

The $B_y (x,t)$ and $E_x (x,t)$ are Fourier transformed over space by $B_y 
(k_n,t) = N_g^{-1} \sum_{j=1}^{N_g} B_y(j\Delta_x,t)\exp{(-ijk_n\Delta_x)}$ 
with $k_n = 2\pi n/N_g \Delta_x$. The $E_x(x,t)$ is transformed accordingly. 
Figure \ref{fig1} displays the time-evolution of the amplitude moduli of 
the dominant $k_1$-mode for $B_y(k,t)$ and of the $k_2$-mode for $E_x(k,t)$.
The $B_y(k_1,t)$ grows exponentially at the rate $\Omega_{i,1}=\Omega_{i,2} 
= 0.25 \, \Omega_p$ until $t\approx 40$ in $L_1,L_2$ and at $\Omega_{i,3} = 
0.22 \, \Omega_p$ until $t\approx 50$ in the $L_3$. The maximum linear 
growth rate obtained from a cold fluid model is $\Omega_{i,f} = \Omega_p 
v_b / c \Gamma (v_b)\approx 0.29 \, \Omega_p$ \cite{Stockem}. The measured 
growth rate is reduced compared to $\Omega_{i,f}$ in particular for the 
simulation $L_3$ by thermal effects, which cause damping at large $k$. The 
amplitude of $B_y(k_1,t)$ saturates and remains almost constant in all 
simulations after $t=80$. The saturation amplitudes are $B_y (k_1,t>80,L_1) 
\approx 0.06$, $B_y (k_1,t>80,L_2) \approx 0.045$ and $B_y (k_1,t>80,L_3) 
\approx 0.02$.
The magnetic trapping mechanism sets in, when $\Omega_{i,j}$ is comparable 
to the magnetic bounce frequency $\Omega_B={( ek_1 v_b B_y/ m_e )}^{1/2}$ 
in physical units \cite{Davidson}. The $\Omega_B$ for the measured growth 
rates above are $\Omega_B / \Omega_p = 0.2$ for $L_1$ and $0.21$ for $L_2$ 
and $L_3$. The agreement is excellent for the simulation $L_3$, but the 
increase of $\Omega_{i,j}$ in the simulations $L_1,L_2$ compared to that 
in $L_3$ does not change $\Omega_B$. What we find instead is, that $B_y
(k_1,t>80,L_j) \propto L_j$. 

The $E_x (k_2,t)$ grows when $B_y (k_1,t)$ has reached a large amplitude
and the growth rate of $E_x (k_2,t)$ is twice that of $B_y (k_1,t)$, as 
it has previously been reported \cite{Califano,Rowlands}. The $E_x(k_2,t)$ 
in the simulation $L_3$ oscillates between a peak value and zero, whereas 
we find damped oscillations around a steady state value in $L_2,L_1$ after 
$t \approx 50$. The peak amplitude of $E_x$ in the simulation $L_2$ exceeds 
that of $L_3$ by the factor 3 and that in $L_1$ is even stronger (not shown).

\subsection{Simulation $L_1$}

\begin{figure}
\begin{center}
\includegraphics[width=8.2cm]{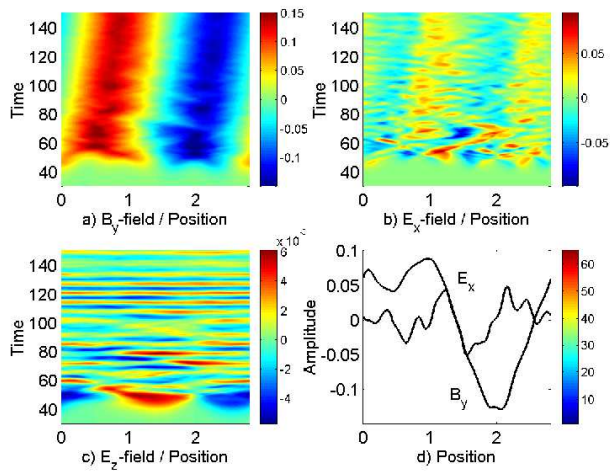}
\caption{The field amplitudes in the box $L_1$: The panels (a-c) show
$B_y$, $E_x$ and $E_z$, respectively. The $B_y$-amplitude reaches a
steady state value, which convects slowly in space. The $E_x$ and $E_z$
components are oscillatory in space and time. The $E_x$ oscillates in
space twice as fast as $B_y$ and the maxima of both are co-moving. The
$E_z$ shows a phase shift of $90^\circ$ compared to $B_y$ when the fields
saturate. The $B_y,E_x$ fields at $t=50$ are displayed by (d) and they
show a correlation only at $x\approx 1.3$.\label{fig2}}
\end{center}
\end{figure}

The evolution of the electromagnetic fields and of the related electron 
phase space is now examined in more detail. Figure \ref{fig2} shows the 
relevant fields in space and time for the simulation $L_1$. It also 
compares the spatial distributions of $B_y(x,t)$ and $E_x(x,t)$ at the 
time $t=50$, when the FI saturates. 
The simulation box fits one oscillation of $B_y (x,t)$ at any fixed time 
$t$ after the saturation, but the more rapid oscillations along $x$ during 
$40<t<50$ indicate that at least initially several modes are competing. 
Eventually, the current channels merge and form a steady state distribution 
in 1D \cite{Davidson,Rowlands}. The magnetic field structure slowly convects 
to larger $x$. The phase speed of this structure is constant after $t=70$ 
and it amounts to $\approx v_{th}/5$. The oscillations of $E_x (x,t)$ are 
more rapid than those of $B_y(x,t)$ for any fixed $50<t<55$. The maxima of 
$E_x (x,t)$ after $t=70$ show a clear correlation with the structure of 
$B_y (x,t)$, because both convect with the same phase speed. The amplitudes 
of $B_y(x,t=50)$ and $E_x (x,t=50)$ show a possible correlation only at 
$x\approx 1.3$. The smaller filaments are merging to the larger one at 
this time, complicating the interpretation of the field correlation. The 
$E_z(x,t)$ has the same spatial oscillation frequency as $B_y (x,t)$ for 
any fixed $40<t<55$ and their phase is shifted by the expected $90^\circ$. 
The amplitude of $E_z$ is significantly lower than that of $E_x$. The 
oscillations of $E_z (x,t)$ after $t=70$ show no pronounced structures.

\begin{figure}
\begin{center}
\includegraphics[width=7.6cm]{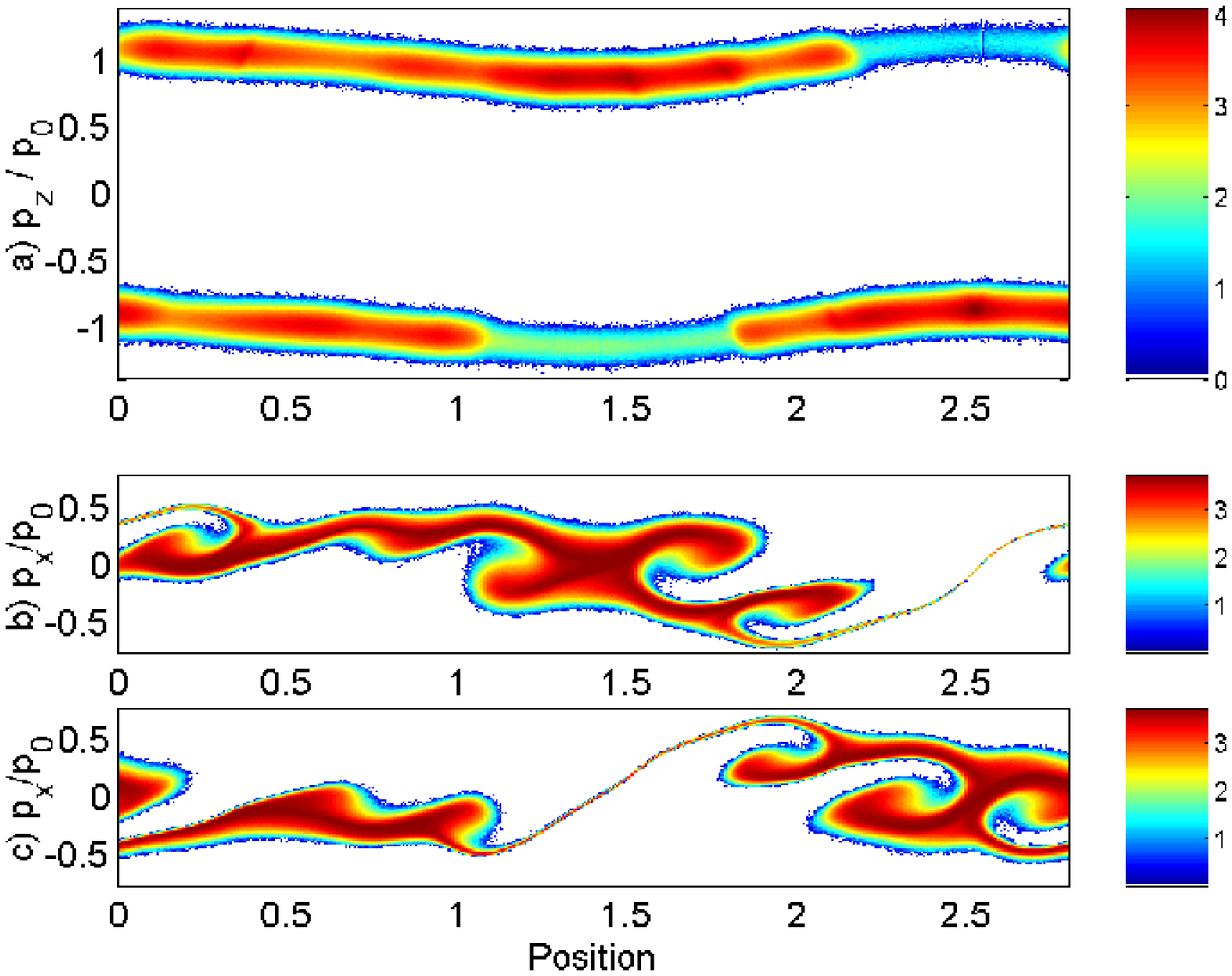}
\includegraphics[width=7.6cm]{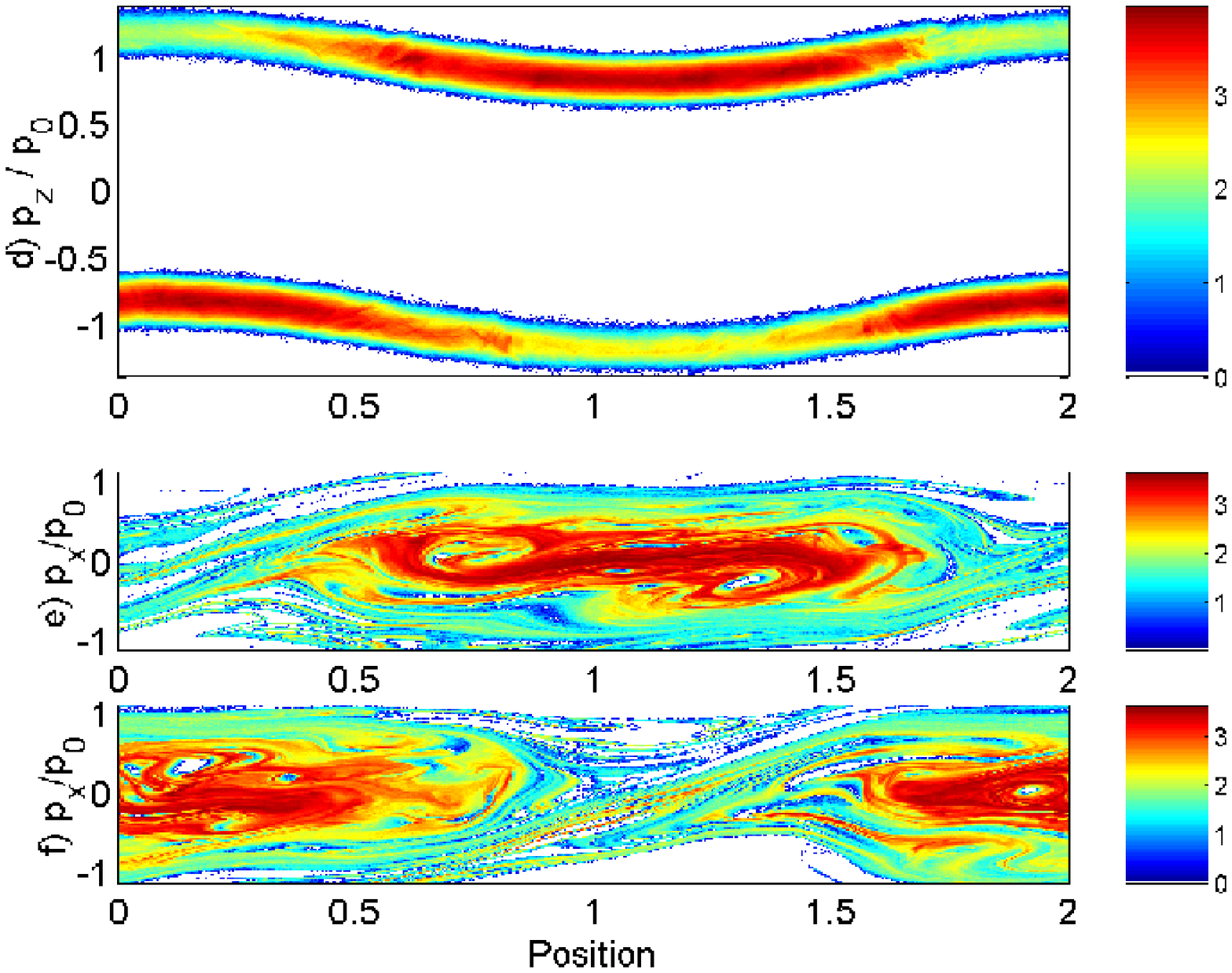}
\caption{The 10-log of the phase space densities in units of CPs at $t=50$ 
(a-c) and $t=120$ (d-f) in the box $L_1$: Panels (a,d) show the $f(x,p_z)$ 
with $p_0 = m_e v_b \Gamma (v_b)$. The beam temperature along $p_z$ is 
unchanged and the $<p_z(x)>_{1,2}$ are obviously modulated in space. The 
density oscillates by the factor $\approx 10^2$. The $f_1(x,p_x)$ is shown 
in (b,e) and $f_2(x,p_x)$ in (c,f). The electrons of both beams spatially 
separate and (e,f) reveal dense turbulent electron clouds immersed in a 
tenuous hot electron background, which reaches a thermal spread $\approx 
p_0$.\label{fig3}}
\end{center}
\end{figure}

Figure \ref{fig3} displays the phase space densities $f_{1,2}(x,p_x)$ and
$f(x,p_z)$ at the times $t=50$ and $t=120$. They show a significant 
modulation and the density maxima of both beams are shifted by $L_1 / 2$ 
at $t=120$. The values of $f_{1,2}(x,p_x)$ and $f(x,p_z)$ vary by two 
orders of magnitude. The mean values $<p_z(x)>_{1,2} = \int p_z f_{1,2}
(x,\mathbf{p}) \, d\mathbf{p}$ for each beam show the same oscillatory 
modulation with $x$ as in Ref. \cite{Rowlands}. The $f(x,p_z)$ at $t=50$ 
also shows that several filaments develop. For example, a density maximum 
is found at $p_x\approx 0.2$ and $p_z = p_0$ that is spatially correlated 
with a minimum at $p_z = -p_0$. The $p_0 = m_e v_b \Gamma (v_b)$. The 
absolute minima of $f_2(x,p_x)$ at $x\approx 1.5$ and of $f_1(x,p_x)$ at 
$x\approx 2.5$ are also not shifted by $L_1/2$. The smaller filaments 
cause the rapid spatial modulation of $B_y (x,t=50)$ in Fig. \ref{fig2}. 
The electrons are heated up in $p_x$ by the saturation of the FI from an 
initial thermal spread of $p_x / p_0 \approx 0.05$ to a peak value of $p_x 
\approx p_0$. The summation of $f(p_x) = \int f(x,\mathbf{p}) \, dx \, dp_y 
\, dp_z$ over many filaments will give a distribution, that decreases 
exponentially over a wide range of $p_x$ \cite{Rowlands}. 

The supplementary movie 1 shows the 10-logarithmic $f_1(x,p_x,t)$ and 
$f_1(x,p_z,t)$ in the simulation $L_1$. It demonstrates that only the 
core electrons in Fig. \ref{fig3} remain spatially confined. This dense 
core population maintains $B_y(x,t>50)$ in Fig. \ref{fig2}. The structures 
in $f(x,p_x)$ in the movie 1, one of which occurs in Fig. \ref{fig3}e) at 
$x\approx 0.7$ and $p_x \approx 0$, resemble phase space holes \cite{Bengt}. 
The potentials of these structures contribute to the $E_x (x,t)$, causing its 
rapid fluctuations in Fig. \ref{fig2}d). These phase space holes complicate 
the identification of the relation between $E_x,B_y$ in Fig. \ref{fig2}d)
in addition to the ongoing merging of small filaments to larger ones.

\subsection{Simulation $L_2$}

\begin{figure}
\begin{center}
\includegraphics[width=8.2cm]{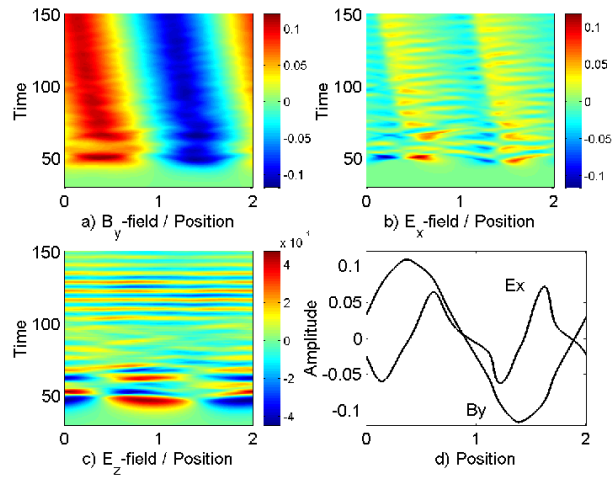}
\caption{The field amplitudes in the box $L_2$: The panels (a-c) show
$B_y$, $E_x$ and $E_z$, respectively. The $B_y$-amplitude reaches a
steady state value, which convects slowly in space. The $E_x$ and $E_z$
components are oscillatory in space and time. The $E_x$ oscillates in
space twice as fast as $B_y$, which is obvious at $t\approx 50$. The 
maxima of both are co-moving. The $E_z$ shows a phase shift of $90^\circ$ 
compared to $B_y$ when the fields saturate. The $B_y,E_x$ fields at $t=50$ 
are displayed by (d) and both show a clear spatial correlation. The $E_x$
is modulated by its first harmonic.\label{fig4}}
\end{center}
\end{figure}

We reduce now the box length from $L_1 = 2.8 \lambda_s$ to $L_2=2\lambda_s$. 
The comparison of Fig. \ref{fig4} with Fig. \ref{fig2} reveals some 
similarities between the respective field evolutions. The structures in
the saturated $B_y (x,t)$ and $E_x (x,t)$ fields are co-moving 
also in the simulation $L_2$ and they have here a phase speed $\approx 
- v_{th}/6$. The $E_z(x,t)$ and $ B_y(x,t)$ have the same spatial oscillation 
period when they saturate at $t\approx 40$, again shifted by $90^\circ$. 
The $B_y(x,t)$ does not show the spatial modulations with $k>k_1$, which 
the $B_y(x,t)$ in Fig. \ref{fig2} does prior to its saturation at $t\approx 
50$. Contrary to the Fig. \ref{fig2}d), the $B_y(x,t=50)$ and the 
$E_x(x,t=50)$ are obviously correlated in Fig. \ref{fig4}d). The $E_x=0$ 
whenever $(B_y \, dB_y / dx) =0$, which suggests the magnetic pressure 
gradient as the origin of $E_x$. The dominant oscillations of $E_x,B_y$ in 
Fig. \ref{fig4}d) are in the $k_2$ and $k_1$ mode, respectively. The $E_x$ 
component evidences furthermore the presence of harmonics.

\begin{figure}
\begin{center}
\includegraphics[width=7.6cm]{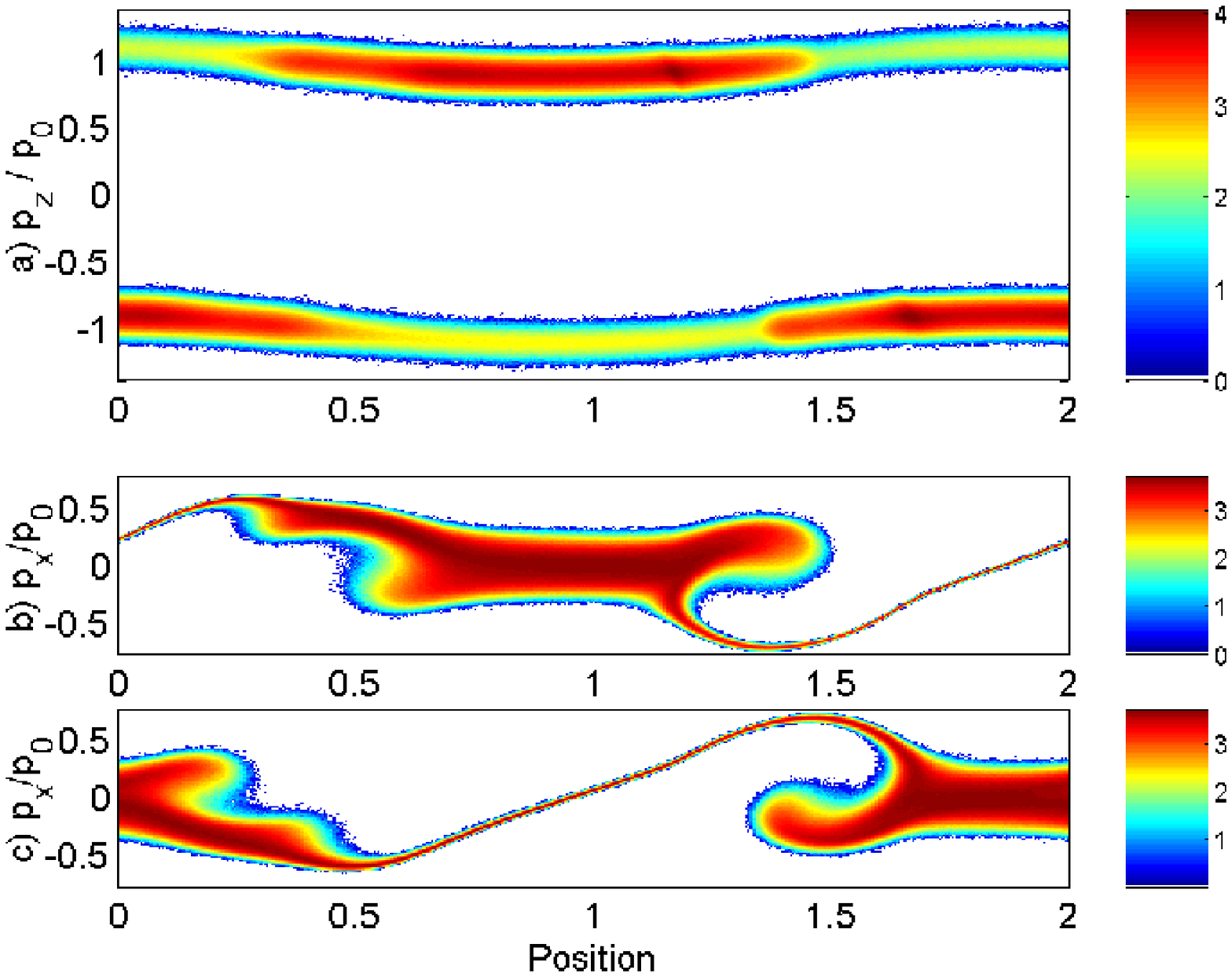}
\includegraphics[width=7.6cm]{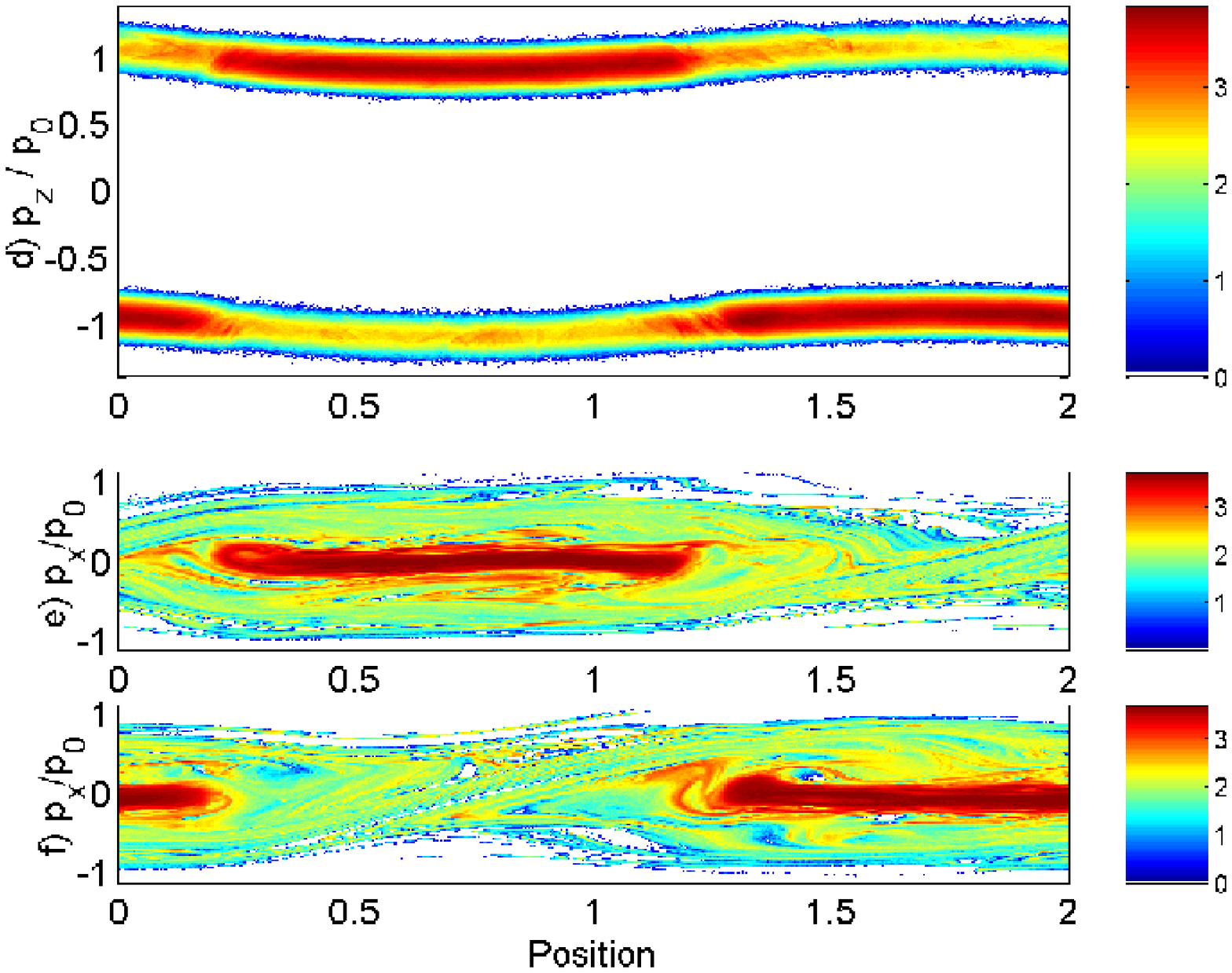}
\caption{The 10-log of the phase space densities in units of CPs at $t=50$ 
(a-c) and $t=120$ (d-f) in the box $L_2$: Panels (a,d) show the $f(x,p_z)$ 
with $p_0 = m_e v_b \Gamma (v_b)$. The beam temperature along $p_z$ is 
unchanged and the spatial oscillations $<p_z(x)>_{1,2}$ of the beams is 
weak. The density oscillates by the factor $\approx 10^2$. The $f_1(x,p_x)$ 
is shown in (b,e) and $f_2(x,p_x)$ in (c,f). Both beams separate in space 
and (e,f) reveal cool electron clouds immersed in hot electrons with a 
thermal spread $\approx p_0$.\label{fig5}}
\end{center}
\end{figure}

Figure \ref{fig5} illustrates the phase space distributions $f_{1,2}(x,p_x)$ 
and $f(x,p_z)$ at the times $t=50$ and $t=120$ in the simulation $L_2$. 
The density maxima are shifted by $L_2/2$ at both times and no further 
density maxima occur. The phase space structures of both beams at $t=50$ 
reveal a high degree of symmetry, evidencing a dominant single filament. 
A second, independently developing filament would break such a symmetry
as in the simulation $L_1$. This is in line with the growth of $B_y (x,t)$ 
in Fig. \ref{fig4} that shows only a modulation with the wavenumber $k_1$. 
The constant slope of $E_x (0.8 < x < 1.1,t=50)$ in Fig. \ref{fig4}d) 
corresponds to a spatially uniform distribution of $f(x,p_x)$ in Fig. 
\ref{fig5}b). The harmonics of $E_x$ in Fig. \ref{fig4}d) are related to 
the phase space structures at the filament boundaries. The $f(x,p_x)$ at 
$t=120$ shows a heated electron population similar to that in Fig. 
\ref{fig3}. The dense electron component in $L_2$ is, however, cooler and 
it shows no vortex structures. The core populations of both beams in Fig. 
\ref{fig5}e,f) are not overlapping, as the current filaments do in the Fig. 
\ref{fig3}.

The supplementary movie 2 animates in time the 10-logarithmic phase space
distributions $f_1(x,p_x)$ and $f_1(x,p_z)$ of the beam 1 in the simulation
$L_2$. The formation of the filaments is demonstrated. The phase space 
evolution shows that the distribution $f(x,p_x)$ contains fewer vortices 
and that the vortices are spread out over a smaller interval of $p_x$ than 
in the simulation $L_1$. The spatial modulation of $<p_z(x)>_1$ of the beam 1 
is also less pronounced. The plasma thus appears to be less turbulent than 
that in $L_1$, which may explain the more obvious relation between $B_y$ 
and $E_x$ in Fig. \ref{fig4} compared to the Fig. \ref{fig2}. The spatial
width of the plasmon containing the dense bulk of the electrons of 
$f_1(x,p_x)$ oscillates in time. The overlap of the filaments in Fig.
\ref{fig5}e,f) is thus time dependent and related to the oscillating
electrostatic field in Fig. \ref{fig4}b).

\subsection{Simulation $L_3$}

\begin{figure}
\begin{center}
\includegraphics[width=8.2cm]{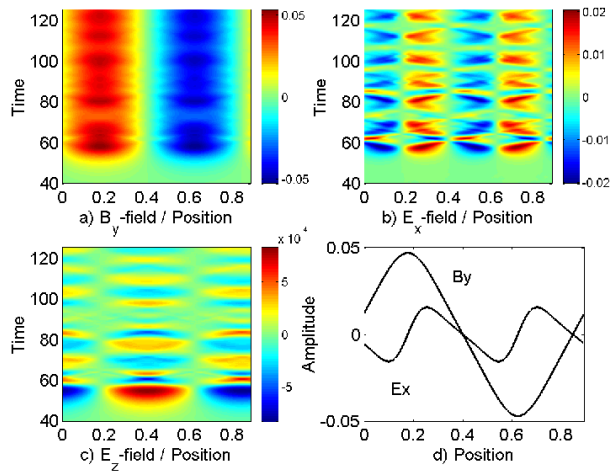}
\caption{The field amplitudes in the box $L_3$: The panels (a-c) show
$B_y$, $E_x$ and $E_z$, respectively. The $B_y$-amplitude reaches a
steady state. The $E_x$ and $E_z$ components are oscillatory in space 
and time. The $E_x$ oscillates in space twice as fast as $B_y$ and
both correlate well for $55<t<125$. The $E_z$ shows a phase shift 
of $90^\circ$ compared to $B_y$ and its amplitude decreases in time. The 
$B_y,E_x$ fields at $t=56$ are displayed by (d) and we find that $E_x=0$ 
when $B_y dB_y / dx = 0$.\label{fig6}}
\end{center}
\end{figure}

The turbulence is reduced further, by decreasing the box length from 
$L_2=2\lambda_s$ to $L_3=0.89 \lambda_s$. We exploit this to examine 
quantitatively and in more detail the relation between the electric and 
magnetic fields and their effect on the particle trajectories. Figure 
\ref{fig6} displays stationary field structures. The $B_y (x,t)$ oscillates 
with the wavenumber $k_1$ in space. The $E_x$ oscillates with the wavenumber 
$k_2$ and it is practically undamped. Both fields display a persistent 
correlation. The $E_z$ component is damped in time and it is shifted 
by the phase $90^\circ$ relative to $B_y$ when the FI saturates. The
damping of $E_z$ does not visibly influence the $B_y$ and $E_x$, 
suggesting that $E_z$ is driven only during the growth phase of the FI
and that it decouples upon saturation. 
The $E_x,B_y$ fields show an excellent qualitative match between $E_x = 0$ 
and $B_y \, dB_y / dx = 0$ when $t=56$. 

The force of the magnetic pressure gradient on a current is expressed as 
\begin{equation}
\mathbf{J} \times \mathbf{B} = -\nabla \cdot \mathbf{B}^2 / 2.
\end{equation}
The derivatives in the $y$ and the $z$ directions vanish in our 1D geometry, 
$B_y \gg B_x,B_z$ and $J_z \gg J_x,J_y$. The magnetic pressure gradient 
force on the electrons can only be mediated through an electric field force 
along $x$. This electric field for the normalized electron charge $-1$ is
then given by $E_B = - B_y dB_y/dx$. The $E_x (x,t>56)$ oscillates in Fig. 
\ref{fig6} in space and time between $E_x = 0$ and an extremum. The peak 
field moduli $\approx 0.02$ are most suitable for a comparison with $E_B$, 
due to the high signal-to-noise ratio. 

\begin{figure}
\begin{center}
\includegraphics[width=8.2cm]{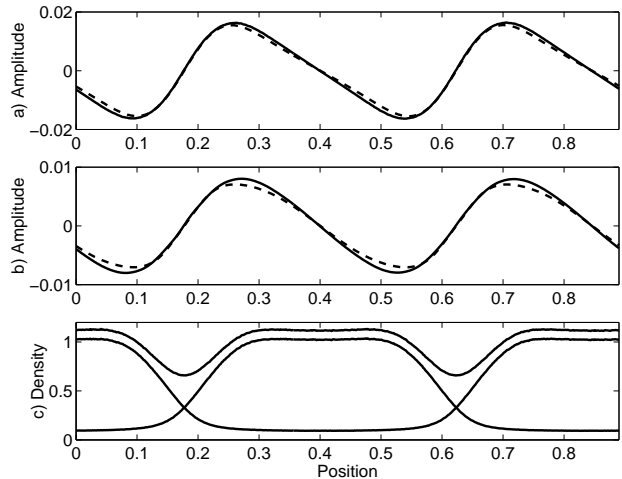}
\caption{(a) The $E_x (x,t=56)$ (dashed curve) and $2E_B$ (solid curve). 
(b) The $\tilde{E}_x (x)$ (dashed line) and $\tilde{E}_B$ (solid line),
which have been averaged over $56<t<125$. (c) The number densities for 
$t=56$, normalized to $2n_e$, of both beams separately (beam 1 is 
almost confined to $0.2 < x < 0.6$) and both densities added together. 
The total beam density is modulated by about $30\%$. The density of the 
individual beams varies by an order of magnitude.\label{fig7}}
\end{center}
\end{figure}

Figure \ref{fig7}a) displays the $E_x (x,t=56)$ when the FI has just 
saturated and compares it with $E_B$. 
It turns out that $E_x (x) \approx 2E_B (x)$ at $t=56$, when the peak
amplitude of $E_x$ is reached. The electric field amplitudes can be 
averaged over the time interval $t_1=56$ to $t_2 = 125$, for which
the field structures do not convect and oscillate around a constant mean 
field, to give $\tilde{E}_x = {(t_2-t_1)}^{-1} \int_{t_1}^{t_2} E_x (x,t) 
\, dt$ and $\tilde{E}_B = {(t_2-t_1)}^{-1} \int_{t_1}^{t_2} E_B (x,t) \, 
dt$. The $\tilde{E}_x$ and $\tilde{E}_B$ match in Fig. \ref{fig7}b). The 
magnetic pressure gradient is thus the origin of the electrostatic field. 
The system is oscillating around the equilibrium because our initial 
conditions did not correspond to a steady state configuration. The same 
oscillations of $E_x$ around a mean field as well as the correlation 
between $B_y$ and $E_x$ can also be observed in Fig. \ref{fig2} and Fig. 
\ref{fig4} at late times for the larger boxes. The $\tilde{E}_B$ correlates 
well with the $\tilde{E}_x$ in the simulations $L_1,L_2$, although the 
curves do not match as accurately as in Fig. \ref{fig7}. This is due to 
the more turbulent plasma and because the convection of the structures 
imposes either constraints on the integration time or requires a 
transformation into a moving frame, the speed of which has to be estimated.

Figure \ref{fig7}c) shows the normalized number density distributions 
$n_{1,2}(x)= {(2n_e)}^{-1} \int f_{1,2}(x,\mathbf{p})\, d\mathbf{p}$ of 
the individual beams and also the summed distribution $n_1(x)+n_2(x)$ at 
$t=56$. The spatial oscillation period of either $n_1(x)$ or $n_2(x)$ is 
$L_3$, while that of the $E_x (x,t=56)$ is $L_3/2$. The phase space 
structures of electron phase space holes in an unmagnetized plasma 
would have the same periodicity as the electrostatic field \cite{Bengt}. 

\begin{figure}
\begin{center}
\includegraphics[width=7.6cm]{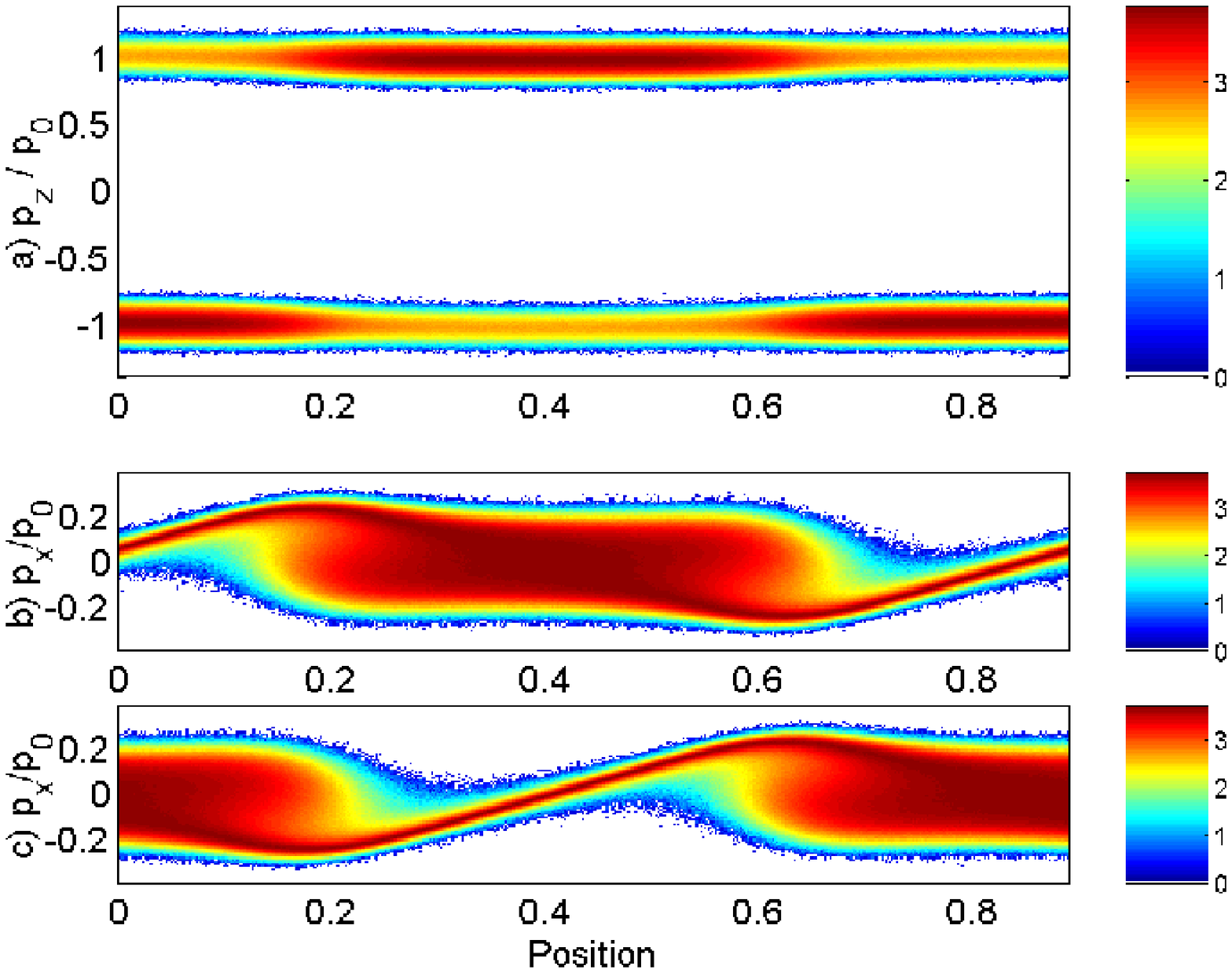}
\includegraphics[width=7.6cm]{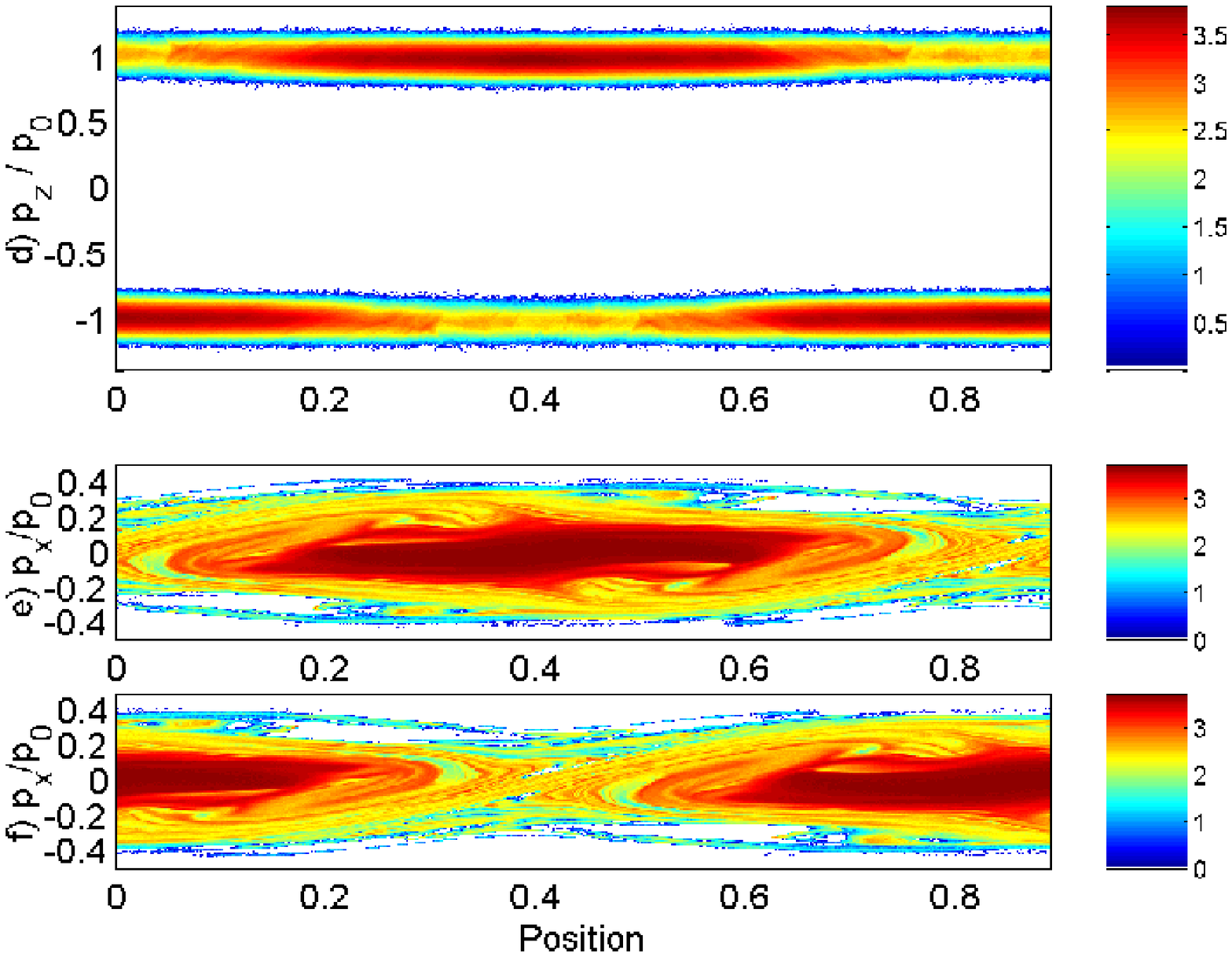}
\caption{The 10-logarithmic phase space densities in units of CPs at $t=56$ 
(a-c) and $t=120$ (d-f) in the box $L_3$: Panels (a,d) show the $f(x,p_z)$ 
with $p_0 = m_e v_b \Gamma (v_b)$. The temperature along $p_z$ and the
$<p_z(x)>_{1,2}$ of each beam are unchanged. The density oscillates by
the factor $\approx 10$. The $f_1(x,p_x)$ are shown in (b,e) and the 
$f_2(x,p_x)$ in (c,f). Both beams separate in space and (e,f) reveal cool 
electron clouds immersed in a tenuous electron population with a thermal 
spread $\approx 0.4 p_0$.\label{fig8}}
\end{center}
\end{figure}

Figure \ref{fig8} shows the phase space distribution of the electrons in 
simulation $L_3$ at the times $t=56,120$.
The $<p_z(x)>_{1,2}$ are practically unchanged along the x-direction. A 
spatial modulation is caused by the $\mathbf{E}\times \mathbf{B}$-drift, 
the speed of which is given by the product of $E_x$ and $B_y$ in our 1D 
simulations. The amplitudes of $B_y$ and of 
$E_x$ both increase according to the Figs. \ref{fig2}, \ref{fig4} and
\ref{fig6} as the filaments get larger and, hence, the drift speed 
contribution to $p_z$.  The amplitudes of $E_x$ in the simulations $L_1,L_2$ 
are several times the one in the simulation $L_3$ and their drift electric
fields $v_b B_y$ are larger. The simulation boxes are also longer. The 
electrostatic potentials in the simulations $L_1,L_2$ thus exceed by far 
that in the simulation $L_3$ and the electrons can reach higher kinetic 
energies. Consequently, the spread in $p_x$ of the phase space distributions 
of both beams in the simulation $L_3$ is less than half of that in the 
simulations $L_1,L_2$ at $t=120$.

The supplementary movie 3 animates in time the evolution of the 10-logarithmic
distributions $f_1(x,p_x)$ and $f_1(x,p_z)$ in the simulation $L_3$. The 
$f_1(x,p_z)$ evidences that the electrons are re-distributed along $x$, but 
not along $p_z$. The electron flow along $x$ oscillates. The $f_1(x,p_x)$ 
after the saturation of the FI has a dense electron core, which rotates in 
the $x,p_x$-plane around $x=0.4$. Two vortices in the dense electron core, 
presumably electron phase space holes, are convected with this rotating flow.

\begin{figure}
\begin{center}
\includegraphics[width=8.2cm]{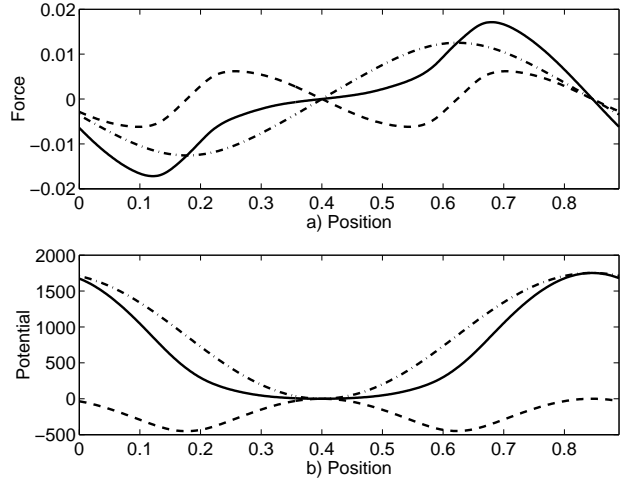}
\caption{The fields $\tilde{E}_j$ and the potentials $\tilde{U}_j$, both 
averaged over the time interval $68<t<125$: a) shows the electric field 
$\tilde{E}_x$ (dashed), the $\tilde{E}_D = v_b \tilde{B}_y$ (dash-dotted) 
and $\tilde{E}_T = \tilde{E}_x+\tilde{E_D}$ (solid line). Positive 
$\tilde{E}_j$ accelerate electrons into the negative $x$-direction. b) 
shows the potential $\tilde{U}_x$ (dashed), $\tilde{U}_D$ (dash-dotted) 
and the $\tilde{U}_T$ (solid). The potential at $x=0.4$ is the reference 
potential.\label{fig9}} 
\end{center}
\end{figure}

All simulations evidence that a core of cool electrons is spatially 
confined. Their circular phase space motion in the movies around the 
equilibrium points $x_e$ with $E_x(x_e) = B_y(x_e) = 0$, for example 
$x_e=0.4$ in the movie 3, furthermore reveals, that they are trapped 
by an electrostatic potential in the $x,p_x$ plane. We consider the 
quasi-equilibrium established in the simulation $L_3$ after the FI
has saturated. We calculate $\tilde{E}_x = {(t_2-t_1)}^{-1} \int_{t_1}^{t_2} 
E_x (x,t) \, dt$ and $\tilde{B}_y = {(t_2-t_1)}^{-1} \int_{t_1}^{t_2} B_y 
(x,t) \, dt$ for the simulation $L_3$. We integrate both fields from
$t_1 = 68$ to $t_2 = 125$, when the equilibrium is established. The
weak modulation of $p_z$ and thus $v_z$ of the CPs in $L_3$ allows us 
to estimate the drift electric field as $\tilde{E}_D = v_b\tilde{B}_y$ 
and the total electric field $\tilde{E}_T = \tilde{E}_x + \tilde{E}_D$. 
The average potentials $\tilde{U}_j(x) = U_{0,j}+\int_0^x \tilde{E}_j
(\tilde{x}) d\tilde{x}$ with the indices $j=x,D,T$ are calculated from 
these fields and $U_{0,j}$ is set such that $\tilde{U}_j(x_e=0.4)=0$. The 
potentials are expressed in Volts, allowing a straightforward comparison 
with the particle kinetic energies. The average fields and potentials
are displayed by Fig. \ref{fig9}.
The electric fields are such that $\tilde{E}_x$ destabilizes the equilibrium
position $x_e = 0.4$, because the negative $E_x (x>x_e)$ close to $x_e$ 
accelerates the electron in the positive direction and the positive 
$E_x (x<x_e)$ close to $x_e$ in the negative direction. The $E_D (x\approx
x_e)$ is confining the electrons around $x\approx x_e$.
The $|E_D|>|E_x|$ for $x\approx x_e$ and $E_T$ is thus a confining force.
However, the electron acceleration at $x \approx x_e$ is decreased and
increased at larger $|x-x_e|$. This is reflected also by the potentials.
The magnetic potential invoked by Ref. \cite{Davidson} is dominant. However,
the electrostatic field flattens the bottom of the potential and steepens
the walls. This modified potential results in a bouncing time of electrons
that differs from that in Ref. \cite{Davidson}. The asymmetry of the potential 
close to the stable equilibrium $x_e \approx 0.4$ of the beam 1 and the 
stable equilibrium $x \approx 0.85$ of the beam 2 arises from the dependence 
of $\tilde{E}_D$ on the beam velocity. The velocity is $v_b$ for beam 1 and 
$-v_b$ for beam 2. The $E_x$, on the other hand, acts on both beams in the 
same way.

The CPs of the beam 1 should follow almost straight paths close to $x_e=0.4$ 
and they should be rapidly reflected for $|x-x_e|>0.2$. The potential 
difference $\Delta_U = max(\tilde{U}_T)-min(\tilde{U}_T) \approx$ 1700 V 
in the simulation $L_3$ should trap electrons with speeds up to $\Delta_v 
={(2 e \Delta_U / m_e)}^{1/2} / v_b \approx 0.27$. This matches the momentum 
spread of the cool core population in movie 3 or Fig. \ref{fig8}. The 
oscillations of $E_x$ in Fig. \ref{fig6} explain the periodic release of 
electrons from this cool core seen in the movie 3 and the oscillatory force 
imposed on the electrons by $E_x$ contributes to their heating.

\begin{figure}
\begin{center}
\includegraphics[width=8.2cm]{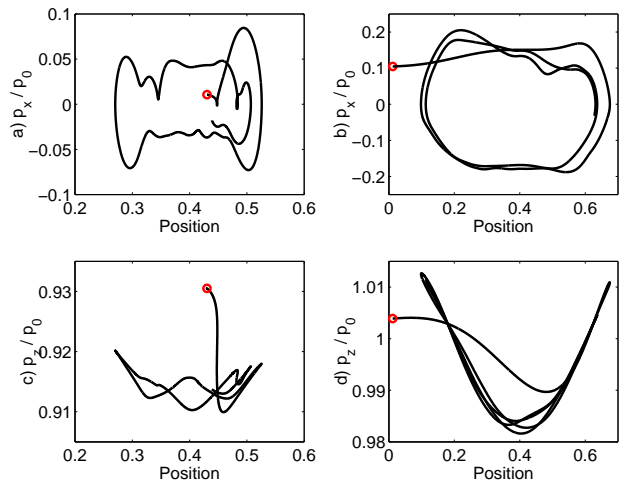}
\caption{The particle trajectories of two selected CPs: (a,c) show the
$x,p_x$ and $x,p_z$ diagrams of the CP 1. (b,d) show the corresponding 
diagrams for the CP 2. The red circle denotes the starting point of the 
trajectory. Both CPs follow straight paths in the $x,p_x$ plane for
$0.33<x<0.47$ and they are rapidly reflected outside this interval. The 
$E_xB_y$ drift imposes diagonal paths of the CP 2 in the $x,p_z$ plane.
\label{fig10}}
\end{center}
\end{figure}

The trajectories of two CPs in the confined structure of beam 1 are 
followed in time in Fig. \ref{fig10} in order to compare their dynamics 
to that expected from $\tilde{E}_T$.
The red circle denotes the time, when the CPs start interacting with the 
fields and the trajectories are followed until $t=125$. The CP 1 has a 
low initial modulus of $p_x$ and CP 2 a high one. Both CPs follow straight 
paths in the interval $0.33 < x < 0.47$, in which $\tilde{E}_T$ in Fig.
\ref{fig9} is small. The phase space path of the faster CP 2 is smoother
than that of CP 1. The low speed of CP 1 implies a long crossing time of 
the interval with a low modulus of $\tilde{E}_T$ and the CP 1 experiences 
several oscillation cycles of $E_x$. The simultaneous action after the
saturation of the FI of the quasi-stationary $B_y$ and the oscillatory 
$E_x$, which both vary in space, implies that the electron acceleration 
is a function of the position and of the phase of $E_x$ relative to $B_y$. 
This phase-dependence results in a randomization of the particle trajectories, 
contributing to the plasma heating. The faster CP 2 crosses the bottom of 
the potential more rapidly and it experiences lower relative speed changes 
by the $\tilde{E}_T$. The particles are reflected outside the interval 
$0.33 < x < 0.47$ by the $|\tilde{E}_T|$. Both CPs are trapped because 
their speed is less than that required to overcome $\Delta_U$.

The $E_z$-fields are weak but not negligible. A comparison of the Figs. 
\ref{fig2}, \ref{fig4} and \ref{fig6} demonstrates, that its time-evolution  
depends on the box length. The $E_z$ grows in all simulations prior to 
the saturation of the FI and its phase is shifted by $90^\circ$ relative to 
$B_y$. It is thus pumped through Ampere's law by the growth of $J_z$. The 
initial, low-frequency oscillations of $E_z$ with $k=k_1$ damp out in all
simulations. They are replaced in the simulations $L_1,L_2$ by faster 
oscillations that have $k=0$ and $k=k_1$. This is not the case for the 
$E_z$ in the simulation $L_3$. The power spectrum of $E_z$ reveals the
origins of the damping of the low-frequency modes and of the growth 
of the high-frequency modes. This power spectrum is obtained by Fourier
transforming $E_z(x,t)\rightarrow E_z(k,\omega)$ over the full box and 
simulation time. The power spectrum $\hat{E}(k,\omega) = {|E_z(k,\omega)|}^2$, 
which we plot for the simulations $L_2,L_3$ in Fig. \ref{fig10}.

\begin{figure}
\begin{center}
\includegraphics[width=8.2cm]{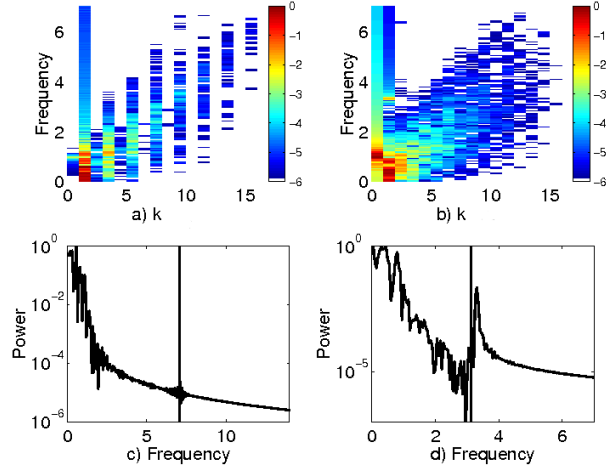}
\caption{The 10-logarithmic power spectra of $E_z$, integrated 
over $0<t<125$ and normalized to their peak value. Frequencies are normalized
to $\Omega_p$. (a,c) correspond to the simulation $L_3$ and the $k$ is
normalized to $2\pi/L_3$. (b,d) correspond to the simulation $L_2$ and
$k$ is normalized to $2\pi/L_2$. (a) reveals a discrete wave spectrum 
in $k$ while the spectrum in (b) is turbulent. The frequency spread is due 
to the growing amplitude of $E_z$. The frequency of the fast X-mode with 
$\omega \approx ck_1$ is indicated by vertical lines in (c,d). Only (d) 
shows a pumping of the fast X-mode.\label{fig11}}
\end{center}
\end{figure}
The $\hat{E}(k,\omega)$ reveals the following for the simulation $L_3$. The 
spectrum of $E_z$ consists of the fundamental mode at $k=k_1$ and of secondary 
waves at $k_j$ with j=1+2n and $n=1,2,.$ . The modes with the $k>k_1$ are not 
harmonics of $k_1$. They can thus not originate from a self-interaction of 
$E_z$ or from a coupling of $E_z$ with $B_y$, which has also $k=k_1$. It
appears that $E_z$ is interacting with the $k_2$ mode of $E_x$ and, 
possibly, also with $B_y$. The power spectrum is spread out in frequency,
because $E_z$ initially grows exponentially and is thus not monochromatic. 
The $E_z$ can couple to the fast extraordinary (X-)mode with 
the same polarization. Its frequency $\omega \approx ck$ in physical units 
at $k=k_1$ is high for the simulation $L_3$ due to the high 
$k_1$ of this simulation and no power is transfered from the low-frequency 
turbulence to the high-frequency fast X-mode. The transient oscillations 
of $E_z$ in the simulation $L_3$ damp out. 

As we go from $L_3$ to $L_2$ the frequency of the fast X-mode at $k=k_1$ 
is lowered by a factor 2.25. Figure \ref{fig11} evidences that now wave 
energy is coupled into the fast X-mode at $k=k_1$. This pumping also 
transfers wave power to the mode with $k=0$. This may be achieved by a 
parametric interaction of the low-frequency mode at $k=k_1$ and the 
high-frequency mode at $k=k_1$. The frequency spread of the initial 
modes implies, that it is not necessary to fulfill exactly the frequency 
and wavenumber conservation of the 3-wave interaction. The fast X-modes 
are linearly undamped, explaining the persistent high-frequency oscillations 
of $E_z$ in the Figs. \ref{fig2} and \ref{fig4}. 

\section{Discussion}

In this work we have examined in detail the saturation of the electron 
beam filamentation instability (FI) with one-dimensional PIC simulations. 
The one-dimensional geometry implies that we can not model the FI beyond 
its saturation, because the filament merging is excluded \cite{Lampe}. 
However, we can readily distinguish between the electrostatic fields with 
their polarization vector along the simulation direction and the
electromagnetic waves. The phase space distribution can also be sampled
with a high statistical accuracy. We have modelled here only mobile 
electrons. The simulation compensates their charge by introducing an 
immobile positive charge background that cancels the electron charge. 

The FI grows magnetic fields out of noise by a spatial separation of the
currents of the, initially uniform, electron beams. A broad wavenumber 
spectrum is destabilized and the amplitudes of the individual filamentation 
modes grow aperiodically and exponentially. The magnetic field amplitude 
can not grow indefinitely. The magnetic amplitude saturates, once the 
bouncing frequency of the electrons in the magnetic potential becomes 
comparable to the growth rate of the instability \cite{Davidson}. More 
recently, it has been pointed out that the electrostatic fields upon 
saturation may not be negligible \cite{Califano}. These electrostatic 
fields have been connected qualitatively to the magnetic pressure gradient 
\cite{Stockem,Rowlands}. The observation, that the saturation of the FI is 
qualitatively unchanged by a beam-aligned magnetic field, has led to the 
suggestion in Ref. \cite{Stockem}, that it is not the magnetic trapping 
but the electrostatic fields that quench the instability. This is, because 
the bouncing frequency of electrons depends on the strength of the 
flow-aligned magnetic field, while the electrostatic field does not; the 
pressure gradient of a uniform magnetic field vanishes and leaves unchanged 
the electrostatic field.

The electron beams in our simulations have been equally dense and cool. 
The beams counter-propagated at a non-relativistic speed orthogonally to 
the simulation direction. Picking a box length comparable to the filament 
size has allowed us to obtain a quasi-monochromatic wave spectrum and one
pair of filaments. We have demonstrated quantitatively, that the 
electrostatic field is indeed driven by the magnetic pressure gradient. 
The initial conditions we have used here do not represent a steady state 
and the saturated plasma oscillates around its equilibrium state. The 
amplitude of the electrostatic field oscillates around the mean value, 
which is expected from the magnetic pressure gradient. We have demonstrated
this for the small filament. Small means here, that no sub-structures can 
form because thermal effects \cite{Califano,Bret2} limit the maximum unstable 
wavenumber. We have quantified the relative importance of the electrostatic 
field and of the magnetic field. The electrostatic field is comparable in 
strength to the drift electric field $v_b B_y$ and both are thus relevant 
for the saturation of the FI and for the selected plasma parameters. The 
electrostatic field oscillates in space twice as fast as the magnetic 
one. Its effect is to reduce the force on the electrons close to the 
equilibrium point of the respective filament and to increase it farther 
away from it. The effective potential obtained from a summation of 
the magnetic pressure-driven electrostatic field and of the drift 
electric field is thus not a cosine. 

The electrons can move almost freely through the potential well and are
reflected at a relatively thin layer, as we have demonstrated for two
representative computational particles. The effective potential is not 
symmetric with respect to both filaments. This is, because the drift 
electric field depends on the beam flow direction, while the magnetic 
pressure-driven electric field acts on all electrons the same way. The 
wall of the potential that is confining one filament is located well 
inside the potential well of the second filament. Filaments thus remain
only separated if they have opposite beam flow directions, through
which they can overlap without merging. This effective potential will 
also accelerate ions \cite{Califano}. This ion acceleration is, however,
exaggerated in 1D simulations where the potential is stationary in
space. The filaments merge in higher-dimensional simulations
and the potentials are not longer stationary. We have thus deliberately 
omitted to consider ion effects.

The electrons are heated up when the potential grows and also by their 
simultaneous interaction with the oscillatory electric field and the 
steady magnetic field. A dense and spatially confined electron population 
(plasmon) maintains the current responsible for the magnetic field after 
the heating has taken place. A hot electron population overcomes the 
potential well and presumably contributes to the quenching of the FI,
which can be suppressed by a high temperature orthogonal to the beam
flow direction \cite{Bret2}. 

We have then assessed the impact of the filament size on its dynamics.
The probability distribution for the filament size has been sampled using 
a long one-dimensional simulation box in Ref. \cite{Rowlands}. We have 
examined here two other filaments, the size of which we have selected
according to the probability distribution. The size of the medium filament 
is close to the most common size. The large filament has a size close to 
the maximum observed one. Larger filaments can only form, if filaments 
merge in higher-dimensional systems \cite{Lampe,Dieckmann} because the 
growth rate of the FI decreases, as we go to smaller wavenumbers. Long 
waves are then outgrown by shorter waves. The electrostatic fields of 
the medium and the large filament have been correlated with the magnetic 
field after the FI has saturated and they have oscillated around an 
equilibrium value. This equilibrium value is close to the magnetic pressure 
gradient-driven field, as for the small filament. We have not shown it,
because the fit is not as accurate as for the small filament, due to the 
drift of the structure and the increased levels of turbulence. The
sub-structure, i.e. the merging of smaller filaments in the large 
filament, prior to the saturation of the FI has furthermore prevented
a clear correlation of the electrostatic with the magnetic field at the 
saturation time. Both the medium and large filaments propagated after they 
have saturated, even though the phase speed of the waves driven by the FI 
is zero. The speed has, however, been less than the initial beam thermal 
speed. A filament can be accelerated by the saturation of the FI. Any 
net momentum of the hot and untrapped electron population must be 
balanced by an oppositely directed net momentum of the trapped electrons.
The electromagnetic fields are tied to the trapped electrons and convect
with the plasmons. The direction of the convection is random. The movie in 
Ref. \cite{Rowlands} shows that different convection speeds of neighboring 
filaments result in their spatial oscillation rather than in a convection.

The mean momentum along the beam flow direction has been spatially
modulated. Our comparative study of three filament sizes has shown that
the magnetic field amplitude and the electrostatic field amplitude both 
increase with the filament size. They modulate the mean velocity of the 
beam through the $\mathbf{E}\times \mathbf{B}$-force. This modulation 
thus increases with the filament size, explaining the observation in 
Ref. \cite{Dieckmann} that trapped electrons reach increasingly higher 
speeds as the filaments merge. 

The magnetic field has varied linearly in space over wide spatial 
intervals in all case studies. A magnetic field amplitude with a 
constant gradient results also in a magnetic pressure-driven electric 
field with a constant gradient. The Fourier transform over space of a 
curve with a constant gradient results in a power-law spectrum. This 
is observed also in longer- and in higher-dimensional simulation boxes 
\cite{Rowlands,Dieckmann,JT}.

We have examined the electric field component along the beam flow 
direction. Its wavenumber, the phase shift relative to the magnetic 
field and that it grows during the linear phase of the FI implies 
that it is driven through Ampere's law. These low-frequency oscillations 
are transient modes and they damp out. However, we have confirmed here 
a previous observation \cite{Califano}, that the fast extraordinary mode 
is pumped by this wave component. We could observe this only for the 
two large filaments. The finite box size introduces a discrete wave 
spectrum. The larger the box length, the lower the frequency of the 
electromagnetic mode. The frequency has been low enough in the large 
simulation boxes to absorb wave energy from the low-frequency turbulence. 
The discrete wave spectrum is, however, a finite box (numerical) effect.

\subsection{Acknowledgments}

The authors acknowledge the financial support by an EPSRC Science 
and Innovation award, by the visiting scientist programme of the
Queen's University Belfast, by Vetenskapsr\aa det and by the 
Deutsche Forschungsgemeinschaft. The Swedish HPC2N computer center
has provided the computer time and support.

\end{document}